\documentclass[prd,aps,twocolumn,preprintnumbers,nofootinbib]{revtex4-2}
\usepackage{amsmath,amssymb,mathrsfs}
\usepackage{bm}
\usepackage{graphicx}
\usepackage{url}

\newcommand{\arcsec}{\hbox{$''$}}

\begin{document}

\title{A dual‑moment phenomenological framework for gravitational, electromagnetic and strong interactions}

\author{Zhiguang Tan}\altaffiliation{Corresponding author: tanzg@ccsu.edu.cn}
\author{Shengjie Wang}
\affiliation{School of Electronic Information and Electrical Engineering, Changsha University, Changsha 410003, P.R.~China}

\author{Hua Zheng}\altaffiliation{Corresponding author: zhengh@snnu.edu.cn}
\affiliation{School of Physics and Information Technology, Shaanxi Normal University, Xi'an 710119, China}

\begin{abstract}
The unification of fundamental interactions remains an outstanding open problem in fundamental physics. Most existing unification schemes rely on extra dimensions, exotic symmetries, or unobserved dynamical degrees of freedom. Here we construct a minimal classical phenomenological framework built on primitive charges and motion‑induced kinetic dual moments. Gravitational, electromagnetic, and strong interactions are coherently formulated within a unified inverse‑square‑law structure. We postulate a universal coupling relation connecting source‑charge couplings and moment‑mediated couplings across these interaction sectors. Although weak‑charge degrees of freedom can be formally accommodated within the formalism, full quantitative treatment of weak‑interaction dynamics is left for future work. The model recovers well‑established low‑energy limits for planetary orbits, atomic systems, and heavy‑quark hadrons. Importantly, it yields distinguishable corrections to light deflection by rapidly spinning compact objects, and three independent experimental schemes are proposed to test these predictions.
\end{abstract}

\maketitle

\section{Introduction}
Unifying gravitation, electromagnetism, the strong and weak interactions constitutes a long‑standing pursuit of fundamental physics. The Standard Model successfully describes three quantum interactions within a gauge‑field paradigm \cite{weinberg1967,salam1968}, while general relativity provides the classical description of gravity \cite{einstein1923}. However, these two pillars rest on incompatible mathematical foundations. Numerous unification attempts, including string theory, super-symmetry and extra‑dimension models, introduce substantial new theoretical structures that have yet to receive experimental confirmation \cite{polchinski2017science,arkani2020nature,weinberg2022qft}. Against this background, we explore an alternative classical phenomenological route based on primitive charges and motion‑induced dual moments, in which gravitational, electromagnetic, and strong interactions admit a unified inverse‑square‑law formulation.

A simpler analogy between gravity and electromagnetism emerged long ago as gravitomagnetism (GEM), developed by Heaviside and later extended in relativistic weak field studies \cite{heaviside1893}. Crucially, conventional GEM is merely a low speed, weak gravity approximation of general relativity and applies exclusively to gravitational phenomena \cite{maheshwari2012,mashhoon2023prd,arneth2025}. It does not extend to electromagnetic, strong or weak forces, and cannot supply a universal structural rule for all interactions. Some phenomenological models treat strong and weak interactions as superpositions of fundamental charge interactions \cite{herrero2025}, yet such frameworks fail to account for the origin of gravitational mass interaction. Similarly, gauge theories treat each interaction with separate symmetry groups without revealing a common classical duality pattern \cite{klebanov2021prd}.

Motivated by the source–dual field duality inherent in electrodynamics, this work advances a conceptually distinct picture: the coexistence of source charge interaction and motion‑triggered source‑moment interaction is not unique to electromagnetism, but represents a universal law governing fundamental interactions. All primitive charges sustain inverse‑square source fields, yielding source‑charge interactions regardless of motion state.
Relative motion creates kinetic moments, which excite supplementary dual transverse fields and produce source‑moment interactions. We introduce the notion of primitive charges associated with each force, derive a unified two‑body interaction formalism, and hypothesize the universal relation \(T=Z/c^2\) linking source charge coupling and source‑moment coupling coefficients. The framework is tested against dynamical systems spanning planetary motion, atomic orbits and heavy-quark hadrons \cite{tan2026}. Most importantly, we propose three independent test schemes: improved torsion‑balance measurement, a novel mass current laboratory experiment, and astronomical light deflection observations, which can be used to verify the unified theory and discriminate it from conventional approximations.

Unlike many unification proposals, our construction invokes no extra dimensions, unobserved particles or fine‑tuned parameters, offering a minimal, testable path toward a common description of fundamental forces.
In the following, we first establish the general duality formalism and primitive charge classification, then construct the unified interaction equations, present multi‑scale dynamical comparisons, discuss theoretical implications, and elaborate the three proposed experimental validation schemes.

\section{Review of Duality in Electromagnetic Interactions}
Electromagnetism constitutes the most thoroughly understood and self‑consistent fundamental interaction, characterized by a distinct source‑motion dual structure. Within classical electrodynamics, duality separates into two core laws governing source charge effects and dynamic dual effects, sharing identical structural forms and differing only through the motion state of electric charges.

First, the inverse square source charge force between point charges in vacuum is the Coulomb force:
\begin{align}
\bm{F}_e=\frac{1}{4\pi\varepsilon_0}\frac{q_1 q_2}{r^2}\hat{\bm{r}}.
\end{align}

Second, moving charges excite dynamic magnetic fields. According to the Biot–Savart law, the magnetic field generated by a moving point charge reads
\begin{align}
\bm{B}=\frac{\mu_0}{4\pi}\frac{q\bm{v}\times \hat{\bm{r}}}{r^2}.
\end{align}

Maxwell’s equations provide a rigorous field‑theoretical interpretation of electromagnetic duality. Electric fields originate from the electric charge content of matter, whereas magnetic fields arise exclusively from moving charges (electric currents). Electric fields reflect the spatial distribution of intrinsic charge, and magnetic fields reflect transport effects associated with charge motion. The two form a strict duality between static sources and dynamic motion, excluding the possibility of isolated magnetic monopoles in classical physics.

\section{Primitive Charges and Charge Moments Generated by Motion}
The essence of electromagnetic duality extends beyond the source‑charge force and dual force duality at the field level. At a deeper physical level, motion of any primitive charge in space necessarily generates a corresponding dynamical moment. Such charge moments serve as the fundamental sources exciting magnetic‑like dual fields, potentials and coupling effects.

It must be emphasized that a primitive charge always produces source‑charge inverse square fields irrespective of its velocity. Charge moments and their induced dual magnetic‑like fields are only generated when the primitive charge undergoes relative motion. Static primitive charges possess zero charge‑moment.

We assume that each elementary property of matter corresponds to a quantifiable primitive charge. The four interactions are associated with mass charge $Q_m$, electric charge $Q_e$, color charge $Q_c$ and weak charge $Q_w$, respectively. Different primitive charges possess distinct varieties. Mass charge exists only in one species (positive mass). Electric charge has two species (positive and negative). Color charge has three varieties, and weak charge admits four varieties. To achieve unified quantitative representation, we employ unit vectors on the unit circle to represent unit primitive charges of each species.

\begin{figure}[htb]
\centering
\includegraphics[width=0.35\linewidth]{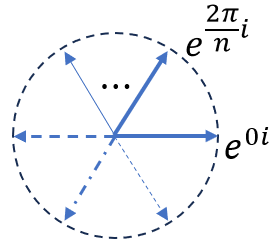}
\caption{Classification diagram of primitive charges.}\label{charg1}
\end{figure}

All unit charges of a given type are uniformly distributed around the unit circle. The opposite vector corresponds to the corresponding anti‑charge. Only primitive charges belonging to the same category can be added or subtracted algebraically. A general primitive charge is expressed as
\begin{align}
Q=\sum_{k=0}^{n-1} \alpha_k e^{2\pi i k/n}.
\end{align}

\begin{table}[htb]
\centering
\caption{Comparison of fundamental interaction charge properties}\label{tab:charge_compare}
\begin{tabular}{l l l l l}
\hline
Property & Mass & Electric  & Color  & Weak \\
\hline
Number of species & 1 & 2 & 3 & 4  \\
Unit charge & $1$ & $1,-1$ & $1,-\tfrac12\pm\tfrac{\sqrt{3}}{2}i\quad $ & $\pm1,\pm i$ \\
\hline
\end{tabular}
\footnotetext{Note: The unit charges listed represent the generators of corresponding gauge symmetry groups.}
\end{table}

The charge moment generated when a primitive charge moves relative to a reference point is defined as
\begin{align}
\bm{P}=\bm{r}\times Q\bm{v}.
\end{align}

Evidently, the mass charge moment corresponds to the familiar angular momentum $\bm{L}=\bm{r}\times m\bm{v}$. The electric charge moment $\bm{P}_e=\bm{r}\times Q_e\bm{v}$ is closely related to magnetic dipole moments. Consider a uniformly charged sphere with total charge $Q$ rotating about its symmetry axis with angular velocity $\omega$. The electric charge moment is
\begin{align}
P_e=\int_V r\,dq\,v=\omega\int_V r^2 dq=\frac{2}{5}Q\omega R^2.
\label{ePe}
\end{align}

If the sphere represents an electron, its spin magnetic moment is given quantum mechanically by
\begin{align}
\bm{\mu}_e=-g_e\frac{e}{2m_e}\bm{S}.
\label{emue}
\end{align}

Here $g_e$ denotes the electron spin $g$‑factor, experimentally measured as $g_e=2.0023193 $ \cite{codata2018}, close to 2. The electron rest mass is $m_e=9.109\times 10^{-31}\,\mathrm{kg}$, and the spin angular momentum satisfies $S=\sqrt{s(s+1)}\hbar$ with $s=1/2$. Replacing $\bm{S}$ with the classical rotational angular momentum $L=\tfrac25 m_e \omega R^2$, we obtain
\begin{align}
\mu_e=-g_e\frac{e}{2m_e}\cdot \frac{2}{5}m_e \omega R^2=\frac{2}{5}\omega e R^2,
\end{align}
consistent with our definition of the electric charge moment as shown in Eq.~\eqref{ePe}.

\section{Unified Description of Two‑Body Interactions}
Source‑charge forces between any pair of primitive charges satisfy the inverse square law
\begin{align}
F=Z\frac{Q_1\cdot Q_2}{r^2}.
\end{align}

In this expression, $Z$ is the source‑charge coupling coefficient governing source charge forces. For mass charge, $Z$ equals the Newtonian gravitational constant $G=6.67\times 10^{-11}\,\mathrm{N}\cdot\mathrm{m}^2/\mathrm{kg}^2$. For electric charge, $Z$ corresponds to the electrostatic constant $k=9.0\times 10^9\,\mathrm{N}\cdot\mathrm{m}^2/\mathrm{C}^2$. For color charge, estimates of the source charge coupling coefficient from fits to heavy meson experimental data yield approximately $Z=7.35\times 10^{-26}\,\mathrm{N}\cdot\mathrm{m}^2$ \cite{tan2026}. We emphasize that these coupling constants depend on the adopted normalization scheme for primitive charges. For instance, the normalization adopted in Ref. \cite{tan2026} sets $\alpha_c=1$, leading to a relatively small value of the color‑charge source coupling coefficient.

Charge moments $\bm{P}$ excite transverse magnetic‑like dual fields dual to the source charge field:
\begin{align}
\bm{B}=T\frac{3(\bm{P}\cdot \hat{\bm{r}})\hat{\bm{r}}-\bm{P}}{r^3}.\label{bm}
\end{align}

Here $T$ denotes the charge moment coupling coefficient associated with charge moment dual forces. In electromagnetism, the two coefficients are connected via the speed of light: $T=k/c^2$. Remarkably, the fitted parameters for strong interactions presented in Ref. \cite{tan2026} also approximately satisfy $T=Z/c^2$. We therefore conjecture a universal relation valid for all categories of primitive interactions:
\begin{align}
T=\frac{Z}{c^2}.
\label{eZT}
\end{align}

This universal relation can be tested by dedicated experiments.
A moving primitive charge immersed in a magnetic‑like dual field experiences a charge‑moment dual force
\begin{align}
\bm{F}=Q\bm{v}\times \bm{B}.
\end{align}

Consider a particle with spin charge moment $\bm{P}_1$ placed in the magnetic‑like dual field produced by a second particle with spin charge moment $\bm{P}_2$. The interaction potential is $U=-\bm{P}_1\cdot \bm{B}_2$, and the corresponding force acting on the particle reads
\begin{align*}
\bm{F}_s &= -\nabla U=\nabla\left(\bm{P}_1\cdot \bm{B}_2\right) \\
&= \nabla\left[\frac{T}{r^3}\big(3(\bm{P}_1\cdot\hat{\bm{r}})(\bm{P}_2\cdot\hat{\bm{r}})-\bm{P}_1\cdot\bm{P}_2\big)\right].
\end{align*}

Two limiting configurations are analyzed:
\begin{enumerate}
\item Collinear alignment. The two spin moments point along the line connecting the particles (the $z$ axis). The force exerted on particle 2 by particle 1 takes the form
\begin{align}
F_z=-\frac{6\,T\,\bm{P}_1\cdot \bm{P}_2}{r^4}.
\end{align}
\item Parallel transverse arrangement. The spin moments remain parallel to each other but perpendicular to the interparticle separation vector (along the $y$ axis):
\begin{align}
F_y=\frac{3\,T\,\bm{P}_1\cdot \bm{P}_2}{r^4}.
\end{align}
\end{enumerate}

Consider a binary system consisting of two objects with masses $m,M$ and spin charge moments $\bm{P}_1,\bm{P}_2$. Mutual primitive‑charge interaction causes mass $m$ to orbit $M$. The centripetal force is supplied jointly by the source‑charge force, orbital charge‑moment dual force and spin–spin interaction force. Classical dynamics gives
\begin{align}
Z\frac{Q_1 Q_2}{r^2}+\frac{Z}{c^2}\frac{\bm{P}_1\cdot Q_1 \bm{v}}{r^3}+\frac{Z}{c^2}\frac{3\bm{P}_1\cdot \bm{P}_2}{r^4}=m\frac{v^2}{r}.
\label{uniF}
\end{align}

\section{Numerical Results}
\begin{figure*}[htb]
\centering
\includegraphics[width=0.95\textwidth]{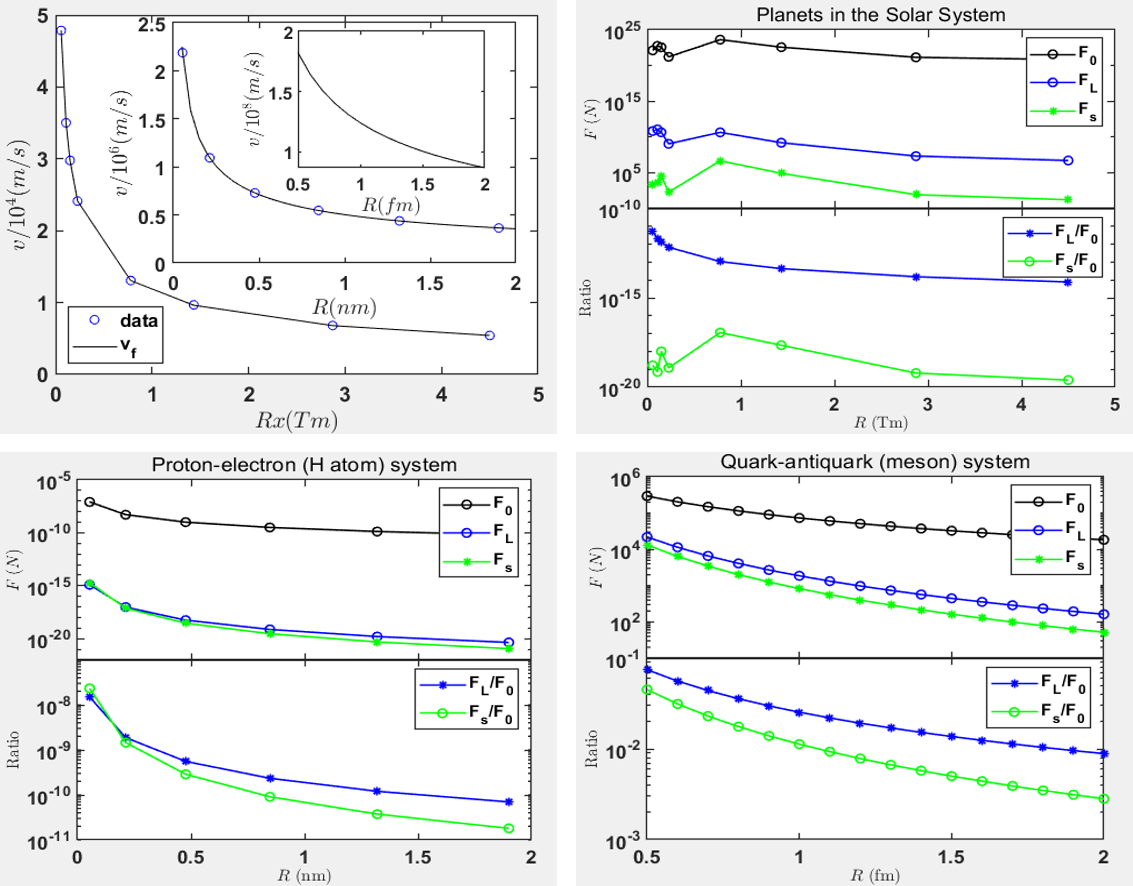}
\caption{Comparison of contributions from different interaction components.}\label{results4}
\end{figure*}

Eq. \eqref{uniF} is applied to investigate dynamical properties of the solar system, hydrogen atoms and meson systems. As shown in Fig. \ref{results4}, the upper‑left subplot compares observed average orbital speeds of the eight planets in the solar system with solutions obtained from Eq.~\eqref{uniF}. Quantum orbital data for hydrogen atoms are also displayed for comparison. For mesons, direct experimental benchmark data are unavailable, and only theoretical predictions are presented. The remaining three subplots in Fig. \ref{results4} illustrate the relative contributions of source charge forces, orbital dual forces and spin interaction forces across different interaction sectors. In all systems, source charge forces dominate the dynamics. Charge moment dual forces act as small corrections.

For gravitational interactions, the ratio between orbital dual forces and source‑charge forces lies in the range $10^{-10}\sim 10^{-13}$. For strong interactions, this ratio increases significantly; at distances below $0.5\,\mathrm{fm}$, the two contributions become comparable in magnitude. The asymptotic freedom observed in QCD can be mimicked classically by competition between static color forces and spin‑dependent color‑magnetic interactions \cite {gross1973,berne2019prd}.

An important observation emerges from the numerical results: all contributions computed above correspond to attractive interactions. Reversing orbital motion direction or switching spin orientation can convert certain attractive contributions into repulsive forces. This classical mechanism reduces the net force acting on orbiting quarks and provides a classical phenomenological analogy for asymptotic freedom observed in the strong‑interaction sector.

\section{Design of Verification Experiments}
To verify the existence of gravitomagnetic gravity, and in particular to validate the universal relation Eq. \eqref{eZT} between the source‑charge gravitational constant $Z$ and the moving‑charge gravitational constant $T$, experimental and observational tests from the following three aspects are proposed to validate our theory.

\begin{enumerate}
\item Improvements to existing gravitational constant measurement apparatus. A schematic diagram of the experimental setup is shown in Fig.~\ref{MF0}. To cancel the contribution of mass induced Newtonian gravity, an identical sphere with mass $M$ can be placed symmetrically with respect to the small test mass $m$. The right panel of Fig.~\ref{MF0} illustrates four distinct configurations for spin alignment. Let $R$ and $r$ denote the radii of the large and small spheres, and $\Omega$ and $\omega$ their respective angular velocities. The resulting spin‑induced gravitomagnetic forces read
\begin{figure}[htb]
\centering
\includegraphics[width=0.9\linewidth]{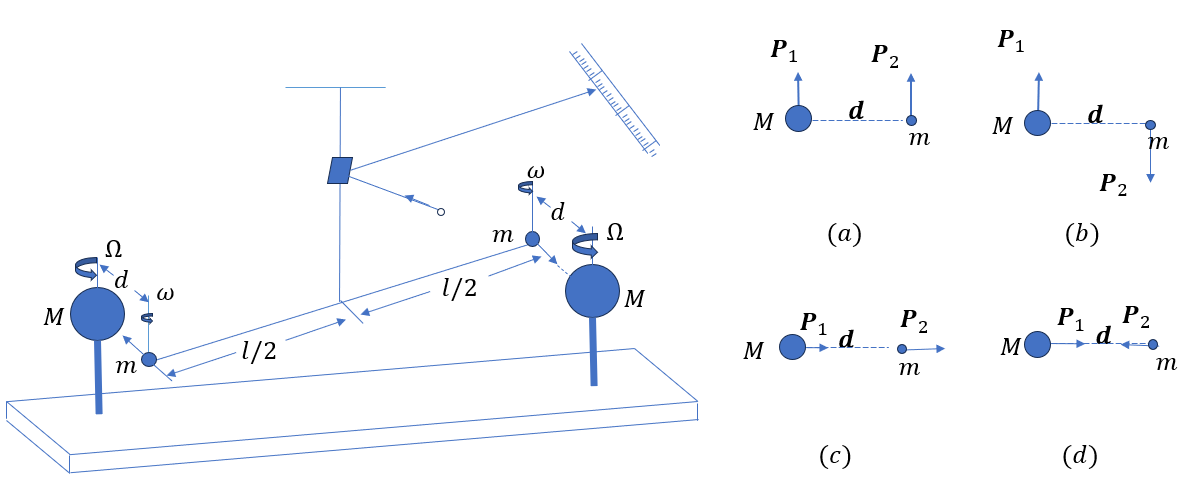}
\caption{Torsion‑balance setup with rotating masses and four spin‑alignment configurations.}\label{MF0}
\end{figure}
\begin{align*}
  (a)\quad F_s &=& \frac{G}{c^2}\frac{3\bm{P}_1\cdot \bm{P}_2}{d^4}=\frac{12GmM\omega\Omega R^2r^2}{25c^2d^4}, \\
  (b)\quad F_s &=&\frac{G}{c^2}\frac{3\bm{P}_1\cdot \bm{P}_2}{d^4}= -\frac{12GmM\omega\Omega R^2r^2}{25c^2d^4}, \\
  (c)\quad F_s &=& \frac{G}{c^2}\frac{6\bm{P}_1\cdot \bm{P}_2}{d^4}=\frac{24GmM\omega\Omega R^2r^2}{25c^2d^4}, \\
  (d)\quad F_s &=& \frac{G}{c^2}\frac{6\bm{P}_1\cdot \bm{P}_2}{d^4}=-\frac{24GmM\omega\Omega R^2r^2}{25c^2d^4}.
\end{align*}

Using typical parameters of torsion balance experiments for estimation:
\(M=1\ \mathrm{kg},\ R=0.05\ \mathrm{m},\ m=0.01\ \mathrm{kg},\ r=0.01\ \mathrm{m},\ d=0.1\ \mathrm{m}\), with rotational speeds \(\Omega=\omega=62.8\ \mathrm{rad/s}\). Substituting into Eqs. (a) and (b), we obtain \(F_s\approx \pm 3.5\times 10^{-29}\,\mathrm{N}\). For Eqs. (c) and (d), \(F_s\approx \pm 7.05\times 10^{-29}\,\mathrm{N}\).

We adopt another set of larger parameters to examine the possible upper bound: let \(M=10\ \mathrm{kg},\ m=0.1\ \mathrm{kg}\), \(R=0.1\,\mathrm{m},\ r=0.02\,\mathrm{m}\), \(d=0.15\,\mathrm{m}\), rotational frequency \(50\ \mathrm{r/s}\), and \(\omega=\Omega=100\pi\approx314\ \mathrm{rad/s}\). Substituting into Eq.~(a) yields \(F_s\approx 2\times 10^{-26}\,\mathrm{N}\), which remains extremely feeble. This value is essentially in agreement with the prediction of general relativity \cite{Thor1988,Bark1975,Ciuf2004}.

\item A novel laboratory experiment is proposed for experimental validation.
The experimental concept is as follows. Two long straight glass tubes connected at their far ends are filled with liquid and arranged horizontally, as illustrated in Fig.~\ref{MFF-fsun5}. A particle emitter is placed at position $A$, on the mid‑line between the two tubes. The liquid flow inside the tubes is controlled by water pumps. When the flow velocity is zero, the particle beam emitted from $A$ travels uniformly to the central point $O$ on the opposite side, subject to equal magnitude and opposite direction Newtonian gravitational attractions from the liquid in the two tubes. If gravitomagnetic forces exist, the particle beam will deflect when the liquid flows. (To eliminate the influence of Earth’s gravity, charged test particles may be employed, with gravitational effects balanced by static electric forces.)

\begin{figure*}[htb]
\centering
\includegraphics[width=0.45\textwidth]{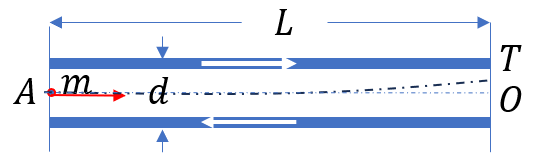}\hfill
\includegraphics[width=0.45\textwidth]{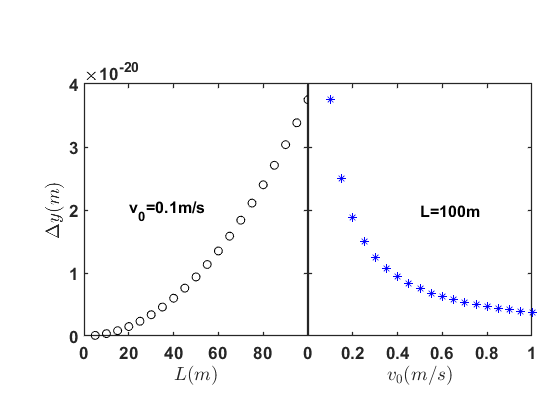}
\caption{Left: Schematic illustration of the particle deflection experiment with counter‑flowing mass‑current tubes. Right: Calculation of gravitomagnetic‑force‑induced beam deflection.}\label{MFF-fsun5}
\end{figure*}

Let $a$ be the inner diameter of each glass tube, $d$ the separation between the two tubes, $L\gg d$ the tube length, $\rho$ the liquid mass density, $V$ the flow speed, $m$ the mass of test particles, and $v_0$ the initial particle velocity. If the gravitomagnetic field described by Eq.~ \eqref{bm} is physical, the deflection of the particle beam under gravitomagnetic force can be calculated.
\begin{enumerate}
\item The gravitomagnetic field generated by a long straight mass‑current tube (mass‑current strength $I_m=\rho s V$) can be obtained by analogy with magnetostatics:
  \begin{align}B_m=\frac{2GI_m}{c^2r}=\frac{4G\rho s V}{c^2d}.\end{align}
\item The net gravitomagnetic force exerted on test particles located on the central axis by two counter‑flowing mass‑current tubes reads
  \begin{align}F_m=2mv_0B_m=2mv_0\frac{4G\rho sV}{c^2d}.\end{align}
\item Subject to this force, particles deviate from the straight line trajectory.
Given the extremely small deflection magnitude, variations of the gravitomagnetic force along the path can be neglected. The transverse displacement of particles is then
  \begin{align} \Delta y=OT=\frac{1}{2}\frac{F_m}{m} t^2=\frac{4G\rho sVL}{c^2dv_0}. \label{delty} \end{align}
\end{enumerate}

Numerical estimation is performed. Take the gravitational constant $G=6.67\times 10^{-11}\,\mathrm{N\cdot m^2/kg^2}$. For water as the working liquid, $\rho=1.0\times 10^3\,\mathrm{kg/m^3}$. The flow velocity driven by pumps is set as $V=2.0\,\mathrm{m/s}$. We choose $d=1\,\mathrm{cm}=0.01\,\mathrm{m}$ and $a=0.4\,\mathrm{m}$, so the cross‑sectional area is $s=\pi (a/2)^2\approx 0.13\,\mathrm{m^2}$. Selected numerical results are plotted in the right panel of Fig.~\ref{MFF-fsun5}. The left sub‑figure shows deflection displacement versus tube length for fixed particle velocity $v_0=0.1\,\mathrm{m/s}$. The right sub‑figure presents deflection as a function of particle velocity for fixed tube length $L=100\,\mathrm{m}$.

It can be seen that the gravitomagnetic interaction is intrinsically extremely weak, since its coupling constant scales as $G/c^2$. From Eq.~\eqref{delty}, the deflection displacement is proportional to the liquid density, tube cross‑sectional area, flow velocity, and the square of tube length, while inversely proportional to tube separation and test‑particle velocity. Observable signals can be enhanced by optimizing these experimental parameters, provided laboratory conditions permit, for example, experiments carried out at heavy‑ion accelerators. Observation of non‑zero beam deflection would serve as experimental evidence for the gravitomagnetic force hypothesis. Precise measurement of deflection data can further constrain or revise the gravitomagnetic coupling coefficient $T$.

\item Validation via astronomical observation of light deflection near stellar surfaces.
Astronomical observations reveal that light rays bend when passing close to high‑density celestial bodies. Einstein interpreted this phenomenon as spacetime curvature within general relativity, an effect independent of celestial body rotation. Within our unified theory framework, this bending can be discussed and computed. In Fig.~\ref{light}, $R$, $M$, and $\omega$ denote the radius, mass, and angular velocity of the celestial body. A light ray coming from infinity grazes the stellar surface. Subject to both Newtonian gravitational attraction and gravitomagnetic effects, the change in photon momentum is given by

\begin{figure}[htb]
\centering
\includegraphics[width=0.8\linewidth]{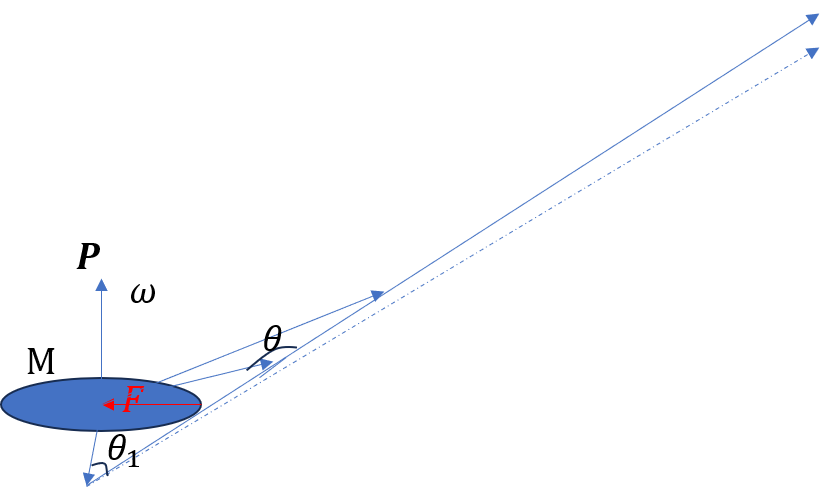}
\caption{Deflection of light grazing the surface of a rapidly rotating celestial body.}\label{light}
\end{figure}

\begin{align}
\Delta p_\bot&=&\int \left[G\frac{Mm}{r^2}\sin\theta+\bm{p}_0\times \left(-\frac{G}{c^2}\frac{\bm{P}}{r^3}\right)\right]dt \nonumber \\
&=&\left(\frac{GMm}{Rc}+\frac{2GM\omega p_0}{5c^3}\right)(\cos\theta_1-\cos\theta_2).
\end{align}

Treating photons as particles with inertial effective mass $m=p_0/c$, and taking $\theta_1=0,\theta_2=\pi$, we obtain
\begin{align}
\Delta \theta=\frac{\Delta p_\bot}{p_0}=\frac{2GM}{Rc^2}+\frac{4GM\omega}{5c^3}. \label{epz}
\end{align}

We note that the first term in Eq.~\eqref{epz} differs by a factor of two from the well‑known general‑relativistic prediction \(\Delta\theta=4GM/(Rc^2)\). Substituting the solar parameters \(M=1.989\times 10^{30}\ \mathrm{kg}\) and \(R=6.957\times 10^{8}\ \mathrm{m}\), the GR formula yields \(\Delta\theta\approx 1.75\arcsec\), consistent with astronomical observations and supporting general relativity \cite{Misn1973,Robe1984}. Our derivation is built upon Newtonian gravity augmented by the relativistic‑mass concept of special relativity. This static‑limit discrepancy arises because our model adopts an inverse‑square source‑charge formalism with motion‑induced kinetic moments, rather than a spacetime‑curvature description; it requires further careful investigation. Nevertheless, the distinct, testable prediction of our framework comes from the rotation‑induced correction term, which can be probed for rapidly‑spinning compact objects independently of constraints from solar static light bending.

The second term in Eq.~\eqref{epz} describes the additional light deflection induced by the rotation of the gravitating body. The ratio of this rotational contribution to the primary static term reads
\begin{align}
\eta=\frac{4GM\omega}{5c^3}\times \frac{Rc^2}{2GM}= \frac{2\omega R}{5c}.\label{epze}
\end{align}

The standard static Schwarzschild solution of general relativity contains no such rotational correction. For a slowly rotating uniform sphere in the weak‑field approximation, the ratio of the first‑order rotational correction from the Kerr metric to the Schwarzschild deflection is \(\eta_{\text{K}}=4\omega R/(5c)\) \cite{Iyer2009}. Thus, the Kerr spin‑induced correction is twice the value obtained from our model. For the Sun, both \(\eta\) and \(\eta_\text{K}\) are of order \(10^{-6}\), which renders them extremely difficult to detect observationally. For rapidly spinning compact objects, however, the equatorial velocity \(v=\omega R\) can be substantial. In such systems, the rotational deflection contribution may reach measurable magnitudes and could be tested by astronomical observations \cite{Chan1983}.
\end{enumerate}

In the Standard Model, weak charge carries clear physical meaning, and weak interactions exhibit point‑like coupling at short distances. According to the unified framework proposed in this paper, moving weak charges should excite weak analogs of gravitomagnetic fields, with coupling constants satisfying $T_w\propto G_w/c^2$. While the primitive‑charge framework can formally accommodate weak charge and weak‑interaction dual‑moment effects, full quantitative computations and symmetry analysis for weak interactions remain work‑in‑progress and are left for future follow‑up studies. Future experiments such as neutral‑current scattering may provide experimental tests for these predictions.

\section{Summary}
This work establishes a minimal phenomenological framework based on primitive charges, including mass charge, electric charge, color charge and a possible weak charge. Gravitational, electromagnetic and strong interactions can be described within a unified mathematical structure. It is anticipated that this framework may be extended to weak-interaction dynamics in an appropriate form in the future. The main conclusions are as follows:
\begin{enumerate}
\item Source‑charge forces associated with primitive charges universally obey the inverse‑square law, acting regardless of whether the primitive charge is static or moving.
\item Motion of primitive charges generates charge moments, which excite magnetic‑like dual fields, leading to charge‑moment dual forces with unified mathematical structure.
\item Source‑charge coupling coefficients and charge‑moment coupling coefficients are universally linked via the speed of light through $T=Z/c^2$.
\end{enumerate}

Electromagnetic theory naturally satisfies these rules. We have proposed three validation routes: torsion‑balance improvement, mass‑current laboratory experiment, and astronomical light‑deflection observation, to search for mass‑magnetic forces.
Astronomical observations offer alternative detection routes. Earlier phenomenological results for color‑magnetic interactions support the framework. The present theory is minimal and self‑consistent, free from redundant hypotheses. It recovers key features of classical gravitation, electromagnetism, and strong interactions and provides a new experimentally accessible route toward unified‑field‑theory research.

\end{document}